\documentclass[useAMS,usenatbib]{mn2e}
\usepackage{epsfig}
\usepackage{amsmath}
\usepackage{multirow}
\usepackage{graphicx}
\usepackage{txfonts}
\begin{document}
   \title[LLGRBs and UHECRs]{Nearby low-luminosity GRBs  as the sources of ultra-high energy cosmic rays revisited}

   \author[Ruo-Yu Liu, Xiang-Yu Wang, and Zi-Gao Dai]{Ruo-Yu Liu$^{1,2}$,Xiang-Yu Wang$^{1,2}$\thanks{E-mail: xywang@nju.edu.cn}, and Zi-Gao Dai$^{1,2}$\\
           $^1$ Department of Astronomy, Nanjing University,
Nanjing, 210093, China\\
$^2$ Key laboratory of Modern
Astronomy and Astrophysics (Nanjing University), Ministry of
Education, Nanjing 210093, China          }

   \date{}

   \pagerange{}\pubyear{}

   \maketitle

   \label{firstpage}

\begin{abstract}
Low-luminosity gamma-ray bursts (GRBs) with luminosity $\lesssim
10^{49} {\rm erg s^{-1}}$ probably consititute a distinct population
from the classic high-luminosity GRBs. They are the most luminous
objects detected so far within $\sim100$ Mpc, the horizon distance
of ultra-high energy cosmic rays (UHECRs), so they are considered to
be candidate sources of UHECRs. It was recently argued that the
energy production rate in UHECRs  is much larger than that in
gamma-ray photons of long GRBs measured by the \textit{Fermi} satellite,
which, if true, would challenge the view that GRBs can be the
sources of UHECRs. We here suggest that many of the low-luminosity
GRBs, due to their low luminosity, can not trigger the current GRB
detectors and hence their contribution to the local gamma-ray energy
production rate is missing. We  find that the real local energy
production rate by low-luminosity GRBs, taking into account the
missing part, which constitutes a dominant fraction of the total
amount, could be sufficient to account for the flux of UHECRs. Due
to the low-luminosity, only intermediate-mass or heavy nuclei can be
accelerated to $\sim10^{20}$ eV. We discuss the acceleration and
survival of these UHE nuclei in low-luminosity GRBs, especially in
those missing low-luminosity GRBs. At last, the accompanying diffuse
neutrino flux from the whole low-luminosity GRB population is
calculated.
\end{abstract}


%

\section{Introduction}
Although the transition energy from galactic to extragalactic origin
in high-energy cosmic ray spectrum remains inconclusive, there is a
general consensus that cosmic rays with energy above the
Greisen-Zatsepin-Kuzmin (GZK) energy of $E\ga E_{\rm
GZK}=4\times10^{19}$ eV,  are of extragalactic origins. These UHE
particles, whether they are protons or heavy nuclei, are attenuated
when they are propagating in the intergalactic space. UHE protons
with energies above the GZK cutoff undergo photopion interactions
with cosmic microwave background (CMB) photons with an attenuation
length of $\sim 100$ Mpc. Coincidentally, heavy nuclei with energy
$\gtrsim 50$EeV will suffer from a strong photo-disintegration
attenuation due to interactions with CMB and cosmic infrared
background (CIB), with an attenuation length $\lesssim 100$ Mpc. As
a result, sources producing UHECRs above $E\ga E_{\rm GZK}$ must be
within $\sim 100$ Mpc. Within this so-called GZK horizon distance,
there are few sources that are powerful enough to be able to
accelerate particles to energies $\sim 10^{20}$ eV. The candidates
include local active galactic nucleus (AGN) jets (e.g.
\citealt{Biermann87,Berezinsky06}), local GRBs
\citep{Waxman95,Waxman04,Vietri95,Wick04,Dermer06,Murase06,Muraseetal06},
and semi-relativistic hypernovae remnants \citep{Wang07a}.  To
accelerate UHECRs, the luminosity of the accelerators must satisfy
the requirement $L\gtrsim 1.5\times10^{42} {\rm erg
s^{-1}}({\Gamma^2}/{\beta})({E}/{10^{20}{\rm eV}})^2 (Z/26)^{-2}$,
where $\Gamma$ and $\beta$ are the bulk Lorentz factor and the
velocity of the shock respectively, $Z$ is the nuclear charge number
of accelerated particles \citep{Waxman05,Farrar09,Lemoine09}. The
brightest sources within this distance are nearby low-luminosity
GRBs (LLGRBs) associated with hypernovae, e.g. GRB980425 with peak
luminosity $L\sim 5\times 10^{46}{\rm erg s^{-1}}$ at a
distance $40$ Mpc (associated with SN1998bw) and GRB060218 with
$L\sim 5\times 10^{46}{\rm erg s^{-1}}$ at distance $140$
Mpc (associated with SN2006aj). They are much dimmer than their
high-luminosity brothers, which are, however, detected at high
redshifts. As a special subclass of GRBs, these nearby
GRBs/hypernovae have been proposed to be candidate sources of UHECRs
\citep{Murase06,Wang07a}. \citet{Wang07b} suggest that
GRB060218-like LLGRBs may arise from the breakout of
semi-relativistic ($\Gamma\sim 2$), radiation-dominated shocks from
the progenitor stars of hypernovae or dense stellar wind surrounding
the progenitor stars. In this scenario, UHECRs can be accelerated at
the forward shock formed when the semi-relativistic ejecta is
expanding in the stellar wind. In the scenario proposed by
\citet{Murase06}, LLGRBs are thought to arise from internal shocks
in relativistic outflow with $\Gamma\sim10$, similar to the case of
high-luminosity GRBs, and UHECRs are accelerated by the same
internal shocks.

GRBs  are discovered preferentially at high redshifts ($z\ga0.5$)
with an isotropic-equivalent energy of $E_\gamma\ga 10^{51-54} {\rm
erg}$, released in a few tenth of seconds to a few tens of seconds.
So far only several nearby GRBs within $\sim 200$ Mpc are detected.
GRB 980425, associated with hypernova 1998bw, is the first-found
peculiar burst detected by \textit{Beppo}SAX and BATSE at a distance
of 38 Mpc, with an isotropic-equivalent total emitted energy of only
$\sim 9.3\times 10^{47}$erg in 1-10000keV band\footnote{hereafter, unless otherwise specified
, the "energy" or "luminosity"  refers to this isotropic-equivalent bolometric
(in 1-10000keV band) one} and a duration of $\sim$35s \citep{Galama98,Kulkarni98,Pian99,Kaneko07}.
GRB 031203,
associated with hypernova 2003lw, was detected by \textit{INTEGRAL}
at $z=0.105$ with an isotropic--equivalent bolometric energy of $\sim 1.7\times
10^{50}$erg and a duration of $\sim$37s \citep{Malesani04,Prochaska04,Kaneko07}.
GRB 060218, associated with
hypernova 2006aj, was detected by \textit{Swift} at a distance of
140 Mpc, with an isotropic-equivalent  energy of $\sim
4.3\times 10^{49}$erg  and a duration of $\sim 2100$s
\citep{Campana06,Mirabal06,Pian06,Kaneko07}. GRB100316D, associated with SN 2010bh, was detected
by \textit{Swift} at distance of $260$ Mpc, with an
isotropic--equivalent bolometric energy larger than $5.9\times
10^{49}$erg and a duration about 1300s \citep{Chornock10,Starling11}.
All these nearby GRBs have low luminosity
and are therefore named low-luminosity GRBs (LLGRBs). Due to their
proximity, the inferred intrinsic rate of LLGRBs is, however, much
higher \citep{Soderberg06a,Liang07,Guetta07,Dai09} than
high-luminosity GRBs (HLGRBs). By checking whether the event rate of
LLGRBs is consistent with  a natural extrapolation of HL GRBs to
low luminosity in a coherent luminosity function (LF), it is found
that LLGRBs likely form an intrinsically distinct population from
HLGRBs (e.g. \citealt{Liang07,Guetta07,Dai09,Virgili09}).

Recently, \citet{Eichler10} find that the energy production rate in
UHECRs above 4EeV is about $10^{2.5}$ times larger than that
contained in gamma-rays recorded   by Fermi from long GRBs. 
{ This appears to be an overestimate, as  later
calculation by \citet{Waxman10} shown that the energy production
rates in gamma-ray photons of GRBs and extragalactic UHECRs are
comparable if some factors are taken into account carefully.  The
discrepancy may be caused by a combination of three factors: 1)
Eichler et al. (2010) assumed that the extra-galactic CR production
rate is about 10 times larger than the UHECR production rate,
arguing that the generation spectrum of the extra-galactic cosmic
rays extends to energy smaller than $10^{18}$ eV (note the spectrum
is $dn/d\varepsilon\propto \varepsilon^{-2.7}$), while the estimate
by Waxman (2010) only includes UHECRs above $10^{19}$ eV; 2)they
assumed different transition energy at which the flux of
extra-galactic cosmic rays dominates over the flux of the Galactic
cosmic rays. Eichler et al. (2010) assumed that extra-galactic
cosmic ray flux dominates over the Galactic component at
$4\times10^{18}$ eV, while Waxman (2010) assumed that it is
dominated  above $10^{19}$ eV; 3) Waxman (2010) also pointed out
that the gamma-ray production rate in Eichler et al. (2010) is
under-estimated, arguing that many distant/faint bursts are missed
by Fermi/GBM. }

Motivated by this problem, we revisit the scenario of LLGRBs as the
accelerators of UHECRs and study whether the energy budget in LLGRBs
is sufficient. Because of the low luminosity of LLGRBs, a lot of
them may not trigger the detector and have therefore eluded
detection. However, these dim GRBs may be still powerful enough to
accelerate UHECRs.

For these LLGRBs, only inter-mediate mass or heavy nuclei can be
accelerated to energies $\sim10^{20}$ eV. The composition of the
observed ultra high energy cosmic rays remains disputed. Recent
observations of the Pierre Auger Observatory (PAO) show a transition
in the maximum shower elongations ${\rm <X_{max}>}$ and in their
fluctuations ${\rm RMS(X_{max})}$ between 5EeV and 10EeV \citep{Abraham10}. These transitions are interpreted as reflecting a
transition in the composition of UHECRs in
this energy range from protons to intermediate mass nuclei. However,
this claim depends on the poorly-understood hadronic interaction
models at such high energies. There is a long-lasting tension
between the spatial UHECR-Active Galactic Nuclei (AGNs) correlation
suggesting protons, and Auger results of $X_{\rm max}$ and $<X_{\rm
max}>$ suggesting a significant heavy element component for UHECR in
the same energy range. Recent update results of PAO show that the
significance of the correlation between UHECRs and nearby
extragalactic matter is decreased \citep{PAO10}, which relives this tension to some extent and allow the
possibility of a heavier composition.

To show whether LLGRBs can be sources of  UHE nuclei, one also needs
to know the survival probability of UHE  intermediate-mass or heavy
nuclei in the sources and the origin of these  nuclei. For the
semi-relativistic hypernova scenario, \citet{Wang08} have shown that
intermediate-mass or heavy nuclei can easily survive in the
hypernova remnant shocks as the shock size is typically very large.
In this scenario, one would also expect a natural origin of
intermediate-mass to heavy nuclei since hypernova remnants are
expanding in  the stellar wind of the progenitor star of hypernovae,
in which intermediate mass nuclei, such as O and C, are enriched.
For the internal shock scenario, \citet{Murase08} discussed the
survival probability of heavy nuclei in LLGRBs for a certain set of
parameter values of the luminosity, shock bulk Lorentz factor and
peak energy of the photon spectrum. However,  these parameters may
vary for different LLGRBs and there are some inherent correlations
among them. We will study their effects on the survival of UHECR
nuclei, especially in those dim LLGRBs that do not trigger the
detector. If these non-triggered LLGRBs are also able to inject UHECRs
into the universe, they will contribute to the energy production
rate in UHECRs, but not to that in gamma-ray photons recorded by
detectors.

The paper is organized as follows. In \S2, we estimate the ratio of
gamma--ray energy production rates between the missing local LLGRBs and
the observable local LLGRBs, as well as the local gamma--ray
and UHECRs energy production rates for two luminosity functions.
We discuss the acceleration and survival process of UHE heavy nuclei
in sources, especially in those LLGRBs that may be missed by the detector in \S 3.
In \S 4, we calculate the accompanying neutrino flux, contributed by
all LLGRBs in the whole luminosity range. We give our conclusion in \S 5.
Throughout the paper, unless otherwise specified, we use eV as the unit
of particle energy and use c.g.s units for other quantities and denote
by $Q_x$ the value of the quantity $Q$ in units of $10^x$.

\section{The local energy production rate by LLGRBs}
To know the local energy production rate by LLGRBs, one needs to
know the luminosity function (LF) of LLGRBs and then sum the
contributions by all LLGRBs over the whole range of luminosity.
Considering that LLGRBs probably form a distinct population from
HLGRBs, here we adopt the LFs of the broken power--law form given by
\citet{Liang07} and \citet{Dai09}. The first one has a form
\citep{Liang07}
\begin{equation}
\frac{dN}{dL}=\rho_0 \Phi_0\left[\left(\frac{L}{L_{\rm b}}\right)^{\alpha_1}
+\left(\frac{L}{L_{\rm b}}\right)^{\alpha_2}\right]^{-1}
\end{equation}
where $\rho_0$ is the local event rate of LLGRBs inferred from the
observed ones, and $\Phi_0$ is a normalization constant to guarantee
the integral over the luminosity function being equal to the local
event rate $\rho_0$. In this luminosity function, the total local
LLGRB rate is insensitive to the minimum luminosity, but fixed by
the break luminosity $L_b$. We take the best fit values for these
parameters from \citet{Liang07} in our following calculation, i.e.
$\rho_0=325{\rm Gpc^{-3}yr^{-1}}$, $L_{\rm b}=10^{47}{\rm
ergs^{-1}}$, $\alpha_1=0$, and $\alpha_2=3.5$.

Another LF for LLGRBs is  suggested as (e.g. \citealt{Dai09})
\begin{equation}
\frac{dN}{dL}=\rho_0\left[\left(\frac{L}{L_{\rm
b}}\right)^{-\alpha_1} +\left(\frac{L}{L_{\rm
b}}\right)^{-\alpha_2}\right],
\end{equation}
which describes HL and LLGRBs together in one form with different
power--law index, joined at the break energy. Here we take
$\rho_0=1.7{\rm Gpc^{-3}yr^ {-1}}$, $L_{\rm b}=5\times 10^{48} {\rm
ergs^{-1}}$, $\alpha_1=2.3$ and $\alpha _2=1.27$, which are
suggested by \citet{Dai09} as the best fit values. In this LF, the
local event rate of LLGRBs is sensitive to the minimum luminosity.
The total event rate of LLGRBs is hard to know since the majority of
LLGRBs at the low-luminosity end could be missed from detection. In
principle, the total event rate of LLGRBs must be lower than the
total rate of type Ib/c supernovae, $\sim 2\times10^4{\rm Gpc^{-3}
yr^{-1}}$ (e.g.  \citealt{Cappellaro99, Dahlen04}). The fraction of
Ib/c supernovae (SNe) which have relativistic outflows is a somewhat
uncertain number. Radio observations of a large sample of  Ib/c SNe
suggest that less than $10\%$ of  Ib/c SNe are associated with GRBs
\citep{Soderberg06a}. We assume two representative values for the
ratio of the local event rate of LLGRBs to that of Type Ib/c SNe,
i.e. $\xi=1\%$ and $\xi=10\%$ in the following calculation. With
these ratios,  one can get the minimum luminosity  $L_{\rm min}$ for
the LF, as shown in Table 1.

With the LFs given above, we can now calculate the local gamma--ray
energy production rates by LLGRBs
\begin{equation}
\dot{W}_{\gamma}(0)=\int_{L_{{\rm min}}}^{L_{{\rm max}}
}E_{\gamma}(L)\frac{dN}{dL}dL,
\end{equation}
where $E_{\gamma}(L)$ is the isotropic equivalent gamma-ray energy
output for  LLGRBs with luminosity $L$. The relation between
$E_{\gamma}$ and $L$ is unknown for LLGRBs due to the small sample
detected so far. We assume $E_{\gamma}\propto L^{k}$, where we take
a wide range for $k$, i.e. $k\in(0,1)$. We take the isotropic
gamma-ray energy in GRB 060218 and GRB 100316D as a reference value,
i.e. $E_{\gamma}=LT_{90}=10^{50}$erg, where $L=10^{47}\rm ergs^{-1}$
and $T_{90}=1000$s. We set the upper limit of integral at $L_{\rm
max}=10^{49} \rm ergs^{-1}$. For the LF in \citet{Liang07}
(hereafter, LFL), there is no constraint on $L_{\rm min}$ since the
total local LLGRB rate is insensitive to the minimum luminosity.
Setting $L_{\rm min}$ to $5\times 10^{45}\rm ergs^{-1}$, which is
sufficiently low to contain all the LLGRBs that have been observed
to date, we have $\dot{W}_{\gamma}(0)\simeq 3.25\times \rm
10^{43}ergMpc^{-3}yr^{-1}$ for $k$=0 and $2.72\times \rm
10^{43}ergMpc^{-3}yr^{-1}$ for $k$=1. The results are comparable for
the case of $L_{\rm min}=5\times 10^{46} \rm ergs^{-1}$, as shown in
Table.~1. On the other hand, for the LF of \citet{Dai09} (hereafter,
LFD), depending on the minimum luminosity $L_{\rm min}$ of the
luminosity function, $\dot{W}_{\gamma}(0)$ is in the range of
$(2-20)\times \rm 10^{43}ergMpc^{-3}yr^{-1}$ for $k$=0 and
$(8.42-14.9)\times \rm 10^{43}ergMpc^{-3}yr^{-1}$ for $k=1$. The
results of $\dot{W}_{\gamma}(0)$ are given in Table 1. These values
are larger than that obtained in \citet{Eichler10} by one to two
orders of magnitude. Furthermore,  the fact that the energy
production rate in the $k=1$ case is less than that in the $k=0$
case in the lower $L_{\rm min}$ case indicates that most gamma--ray
energy production rate is contributed by LLGRBs with relatively
low luminosities (e.g. $L<10^{47}\rm ergs^{-1}$), which may be
difficult to be detected by current GRB detectors.

We now discuss how much the missing LLGRBs contribute to the local
energy production rate. To be detected by \textit{Fermi}/GBM, a GRB
located at a distance $D$ requires to be brighter than a limit
luminosity $L_{\rm lim}=4\pi D^2S_{\rm GBM}/B(e_1,e_2)$, where
$S_{\rm GBM}=1.75\times 10^{-8}\rm ergcm^{-2}s^{-1}$ is the
sensitivity of \textit{Fermi}/GBM in its burst trigger band
(50-300keV, \citealt{vK04,Band08,Imerito08}) and $B(e_1,e_2)$ is the
energy fraction in the detector frequncy window. $B(e_1,e_2)$ is
given by \citep{Bloom01,Imerito08}
\begin{equation}
B(e_1,e_2)=\frac{\int_{e_2}^{e_1}\varepsilon n(\varepsilon)d\varepsilon}
{\int_{E_2}^{E_1} \varepsilon n(\varepsilon)d \varepsilon}
\end{equation}
where $e_1$ and $e_2$ are the upper and lower threshold energy of
the detector, while $E_1$ and $E_2$ are upper and lower limit for
the bolometric gamma--ray spectrum. For \textit{Fermi}/GBM,
$e_1=300$keV and $e_2=50$keV. To be consistent with the inferred
bolometric luminosities of those observed LLGRBs, as mentioned in
Sec.1, we set $E_1=10000$keV and $E_2=1$keV. $n(\varepsilon)$ is the
LLGRB prompt photon spectrum, where $\varepsilon$ is the photon
energy. We assume  a broken power-law spectrum similar to that of
HLGRBs, expressed by $dn/d\varepsilon=n_{\rm b}(\varepsilon
/\varepsilon_{\gamma\rm b})^{-\beta_1}$ for
$\varepsilon<\varepsilon_{\gamma \rm b}$ and $dn/d\varepsilon=
n_{\rm b}(\varepsilon /\varepsilon_{\gamma\rm b})^{-\beta_2}$ for
$\varepsilon>\varepsilon_{\gamma \rm b}$.  We take $\beta_1\simeq 1$
at energies below the break and $\beta_2 \simeq 2$ above the break.
Then the gamma--ray energy production rate by those LLGRBs that are
missed by \textit{Fermi}/GBM can be estimated by
\begin{equation}
\dot{W}_{\gamma, \rm m}=\frac{1}{\frac{4}{3}\pi (D_{\rm
max}^3-D_{\rm min}^3)}\int_{D_{\rm min}}^{D_{\rm max}}\int_{L_{\rm
min}} ^{L_{\rm lim}}4\pi D^2E_{\gamma}(L)\frac{dN}{dL}dLdD,
\end{equation}
where $L_{\rm lim}=2.1\times 10^{46}(D/100{\rm
Mpc})^2/B(e_1,e_2){\rm ergs^{-1}}$. Since 40Mpc is the distance of
GRB 980425, the closest LLGRB ever detected, while 200Mpc is
approximately the attenuation length for iron nuclei of 100EeV, we
take $D_{\rm min}=40$Mpc and $D_{\rm max}=200$Mpc as the lower limit
and the upper limit of the integration respectively. Due to that
$\varepsilon_{\gamma \rm b}$  varies with luminosity, $B(e_1,e_2)$
is a function of the luminosity.  We assume that the Yonetoku
relation holds for LLGRBs as in the case of GRB 060218
\citep{Yonetoku04}, i.e. $\varepsilon_{\gamma \rm b} \propto
L^{1/2}$, and take $\varepsilon_{\gamma \rm b}=10L_{47}^{1/2}$keV.
We define
\begin{equation}
\mathcal{R}=\dot{W}_{\gamma,\rm
m}/(\dot{W}_{\gamma}-\dot{W}_{\gamma,\rm m}),
\end{equation}
as the ratio between the gamma--ray energy production rates in
missing local LLGRBs and that in the observable ones, and present
its values in different cases in Table.~1. We find that the
gamma--ray energy production rate by missing local LLGRBs
constitutes a dominant fraction of the total rate. In some cases, it
is larger than that produced by the observable LLGRBs  by one order
of magnitude. Note that in the $\xi=1\%$ case of LFD, that
$\mathcal{R}$ becomes small is due to an unrealistic high $L_{\rm
min}$, which is higher than the luminosity of detected LLGRBs (e.g.
GRB 980425, GRB 060218). We show how $\mathcal{R}$ changes with $k$
in Fig.~1.

\begin{figure}
\resizebox{\hsize}{!}{\includegraphics{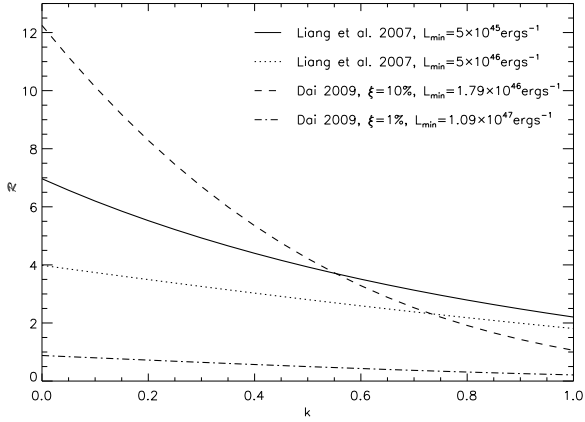}} \caption{The
ratio of the missing to the observable gamma--ray production rate
$\mathcal{R}$ changes with the assumed power--law index  of the
$E_{\gamma}-L$ relation $k$ ($E_{\gamma}=10^{50}L_{47}^k\rm
erg$). Except in the $\xi=1\%$ case for LFD, the missing
gamma--ray energy production rate is several times larger than the
observable one.}\label{fig1}
\end{figure}

\begin{table*}
\begin{center}
\caption{The ratio of gamma--ray energy production rates between the missing local LLGRBs and the observable ones, estimated gamma--ray energy production rate of local LLGRBs and the inferred UHECRs energy production rate in per logarithmic energy interval.
See text for more discussion\label{tbl-1}}
\begin{tabular}{cccccccc}
\hline\hline\\
\multirow{2}{*}{Luminosity Function} & \multirow{2}{*}{$L_{\rm min} (10^{47}$ergs$^{-1}$)} & \multicolumn{2}{c}{$\mathcal{R}^{\mathrm{a}}$} &  \multicolumn{2}{c}{$\dot{W}_{\gamma}(0)^{\mathrm{b}}$($10^{43}$ergMpc$^{-3}$yr$^{-1}$)} &  \multicolumn{2}{c}{$\dot{W}_{\rm UHECR}(0)^{\mathrm{c}}$($10^{43}$ergMpc$^{-3}$yr$^{-1}$)}\\
& & k=0 & k=1 & ~~~~~k=0 & k=1 & ~~~~~k=0 & k=1\\
\hline \\
Liang et al. (2007) & 0.05 & 7.1  & 2.2  & ~~~~~3.25 &  2.72 & ~~~~~2.60 & 2.18 \\\\
Liang et al. (2007) &  0.5 & 4.0  & 1.8  & ~~~~~3.25 &  3.95 & ~~~~~2.60 & 3.16 \\\\
Dai (2009) ($\rho_0=2000$Gpc$^{-3}$yr$^{-1}$) & 0.18 & 12.4 & 1.1 & ~~~~~20.0 & 14.9 & ~~~~~16.0  & 11.9\\\\
Dai (2009) ($\rho_0=200$Gpc$^{-3}$yr$^{-1}$)  & 1.09$^{\mathrm{d}}$ & 0.89 & 0.21 & ~~~~~2.00 & 8.42 & ~~~~~1.60 & 6.74\\\\
\hline
\end{tabular}
\end{center}
\begin{list}{}{}
\item[$^{\mathrm{a}}$]: The ratio of gamma--ray energy production rates between the missing local LLGRBs and the observable ones.\\
\item[$^{\mathrm{b}}$]: The total (the missing $+$ the observable) gamma--ray energy production rate of local LLGRBs.\\
\item[$^{\mathrm{c}}$]: The inferred local UHECRs energy production rate in per logarithmic energy interval, see Eq.~(\ref{WUHECR})
\item[$^{\mathrm{d}}$]: Since the luminosities of GRB 980425 and GRB 060218 are  less than this value by a factor of 2--4, this case may be largely different from the reality. We just show them here for reference.
\end{list}
\end{table*}

Although the obtained gamma-ray energy production rate is highly
dependent on the typical value of $E_{\gamma}$ used (i.e. we used
$E_{\gamma}=10^{50}$erg for a LLGRB with luminosity of $10^{47}\rm
ergs^{-1}$),  which could have large uncertainty, the ratio
$\mathcal{R}$  is only dependent on $B(e_1,e_2)$, or to be more
accurate, the assumed photon spectrum. We also check some other
common combinations of $\beta_1$, $\beta_2$ and $\varepsilon_{\gamma
\rm b}$ and present the results in Fig.~2. In the region that
$L<10^{47}\rm ergs^{-1}$, where the missing LLGRBs reside, the value
$B(e_1,e_2)$ is generally smaller than 0.3. It is mainly because
that the energy window of the detector is so narrow that the
spectral peaks for LLGRBs are typically outside of the window.

\begin{figure}
\resizebox{\hsize}{!}{\includegraphics{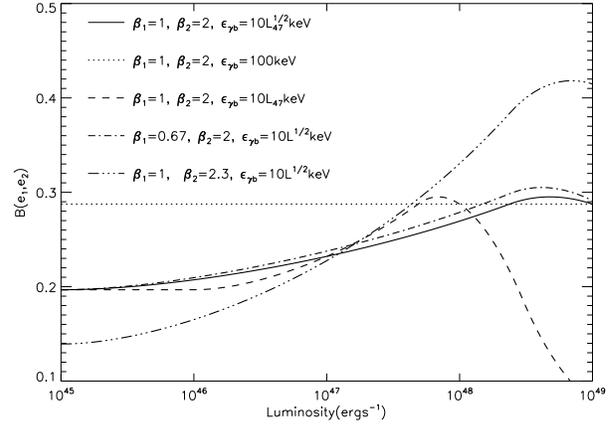}}
\caption{Relation between bandpass ratio $B(e_1,e_2)$ and luminosity with different combinations of parameters of photon spectrum. See text for more discussion.}\label{fig2}
\end{figure}

Recently, \citet{Eichler11} have shown that the contribution of dim or
undetected GRBs to the total all--sky GRB energy flux is small,
based on that the cumulative fluence plateaus at a fluence of
$\sim10^{-5}{\rm erg cm^{-2}}$, which is much higher than the
minimum fluence for both BATSE and \textit{Swift}.
However, we note that the trigger threshold for \textit{Swift}/BAT and
\textit{Fermi}/GBM is about $10^{-8}{\rm erg cm^{-2} s^{-1}}$ and LLGRBs
typical have long duration of $T\sim 10^3 {\rm s}$, so LLGRBs with
fluence lower than $\sim10^{-5}{\rm erg cm^{-2}}$ would not trigger
the detector. Therefore, the plateau only reflects that those GRBs
at the low luminosity end of the HLGRB population contribute little
to the cumulative fluence (which is actually consistent with the LFs
of HLGRBs), irrespective of the LLGRB population.
As a distinctly separate class of bursts at
very low flux levels, LLGRBs could make a significant contribute to
the  cumulative fluence only if the detector is significantly more
sensitive than \textit{Swift}/BAT and \textit{Fermi}/GBM.

In the internal shock scenario for the gamma-ray emission of LLGRBs,
UHECRs are accelerated by the same shocks. So one can estimate the
energy production rate in UHECRs, assuming some equipartition
factors for electrons and protons (nuclei). Assuming a flat spectrum
${d\dot{n}}/{d\varepsilon}\propto \varepsilon^{-2}$ for UHECRs, we
can now estimate the local UHECRs energy production rate per
logarithmic energy interval by LLGRBs
\begin{equation}\label{WUHECR}
\begin{split}
\dot{W}_{{\rm UHECR}}(0)\equiv \varepsilon^2 \frac{d\dot{n}}{d\varepsilon}|_{\rm UHECR}&=\frac{1}
{{\rm ln}(\varepsilon_{A,{\rm max}}/\varepsilon_{A,{\rm min}})}
\dot{W}_{{\rm CR}}(0)\\
&=\frac{1} {{\rm ln}(\varepsilon_{A,{\rm
max}}/\varepsilon_{A,{\rm min}})}
\frac{\epsilon_A}{\epsilon_e}\dot{W}_{\gamma}(0)
\end{split}
\end{equation}
where $\dot{W}_{CR}(0)$ is the local energy production rate for
cosmic ray over the whole energy range, $\varepsilon_{A,{\rm min}}$
and $\varepsilon_{A,{\rm max}}$ are, respectively, the minimum and
maximum energies of cosmic rays. Though the value of
$\varepsilon_{A,{\rm min}}$ is unknown, it is usually assumed that
${\rm ln}(\varepsilon_{A,{\rm max}}/\varepsilon_{A,{\rm min}})\simeq
10$.  Here $\epsilon_e$ and $\epsilon_A$ are, respectively, the
equipartition factors for electrons and protons (nuclei). We adopt
the usual values, $\epsilon_e=\epsilon_B=0.1$ and
$\epsilon_A=1-\epsilon_e-\epsilon_B$ ($\epsilon_B$ is the
equipartition factor for the magnetic field), in the following
calculation. The results of $\varepsilon^2
\frac{d\dot{n}}{d\varepsilon}|_{\rm UHECR}$ are given in Table~1 for
the two LFs considered.  These values are not far from the measured
value of the local energy production rate of UHECRs, $(5\pm2)\times
10^{43}$erg Mpc$^{-3}$yr$^{-1}$ (e.g. \citealt{Katz09,Waxman10}), so
we conclude that local LLGRBs remain viable sources for UHECRs in
terms of the energy production rate.

In the semi-relativistic hypernova scenario for the gamma-ray
emission of LLGRBs, UHECR acceleration occurs in the hypernova
remnant shock region, which is different from the gamma-ray
production region. Nevertheless, the energy production rate in
gamma-rays is still a useful indicator, as it reflects a minimum
energy production rate in the kinetic energy of the
semi-relativistic ejecta, from which the energy of UHECRs is tapped
ultimately.

\section{UHE nuclei from LLGRBs}

\subsection{Acceleration of UHE nuclei}
Now we consider the acceleration of nuclei by semi-relativistic
hypernova remnant shocks or internal shocks in relativistic shells.
For Fermi shock acceleration, the characteristic acceleration
timescale is $t_{\rm acc}=\lambda c/\beta_{\rm sh}^2$, where
$\lambda=\eta R_L$ is the scattering mean free path, $R_L$ is the
Larmor radius of particles and $\beta_{{\rm sh}} c$ is the shock
speed. Here $\eta \sim$ a few, describes the ratio between the
acceleration time and Larmor time ($\eta=1$ corresponds to the
efficient  Bohm diffusive shock acceleration). For a nucleus of mass
$Am_p$ and charge $Ze$, the time needed to accelerate it to the
observed energy of $\varepsilon_{A}$ in the comoving frame is
$t_{{\rm acc}}=\eta \varepsilon_{A}/(ZeB\Gamma\beta_{{\rm sh}}
^2c)$. Cooling of particles restricts the maximum energy. Adiabatic
expansion and synchrotron emission are two important cooling
mechanisms. The cooling time of the former one is given by $t_{\rm
ad}\simeq t_{{\rm dyn}}\simeq R_{{\rm sh}} /\Gamma\beta_{{\rm sh}}
c$, where $R_{\rm sh}$ is the radius of shock acceleration site and
$\Gamma$ is the bulk Lorentz factor of the shock,  while the latter
one is $t_{{\rm syn}}=\frac{6\pi
m_Ac}{(Z^4\sigma_Tm_e^2/m_A^2)\gamma_A B^2}$, where $\sigma_T$ is
Thomson scattering cross section for electrons and
$\sigma_T(\frac{m_e^2}{m_A^2})Z^4$ is the corresponding cross
section for nucleus with $Ze$ charge and $Am_p$ mass.

\subsubsection{Semi-relativistic hypernova  scenario}
Whether particles  can be accelerated to ultra-high energies in the
hypernova scenario has been discussed in \citet{Wang08}, in
which a distribution of the ejecta energy with ejecta velocity has
been assumed. Here for simplicity we only focus on the
semi-relativistic part of the ejecta, whose velocity is
$\Gamma\beta\ga0.5$. According to the simulation of hypernova
explosion such as SN1998bw by \citet{MacFadyen01},  the
isotropic-equivalent kinetic energy of ejecta is roughly constant at
angles larger than $20^{\circ}$, so we assume a spherical  hypernova
ejecta   expanding  into the circum-stellar wind medium. Particles
are accelerated in the region where the ejecta is freely expanding
before being decelerated by the swept-up circum-stellar medium.  The
size of this free-expansion phase region for ejecta of a particular
velocity $\beta_{\rm sh}c$ and kinetic energy $E_K$ is
\begin{equation}
R_{\rm HN}\simeq4\times10^{17} E_{k,51} ({\Gamma\beta_{\rm sh}})^{-2}
\dot{M}_{-5}^{1/2} v_{w,3}^{-1/2} {\rm cm},
\end{equation}
which is much larger than the radius at which gamma-ray photons of
LLGRBs are produced \citep{Wang07b}, where
${\dot{M}}=10^{-5}\dot{M}_{-5}{\rm M_\odot yr^{-1}}$ is the wind
mass loss rate, whose average value is $3\times10^{-5} {\rm M_\odot
yr^{-1}}$ for WR stars, and $v_w = 10^3 v_{w,3}$~kms$^{-1}$ is the wind
velocity \citep{Willis91, Chevalier99}.  During the free expansion phase, the magnetic field
energy density is $B^2/8\pi=2\epsilon_B \rho_w(R_{\rm HE}) c^2
\beta_{\rm sh}^2$, where $\epsilon_B=0.1\epsilon_{B,-1}$ is the
fraction of the equipartition value of the magnetic field energy and
$\rho_w$ is the mass density of the stellar wind at radius $R_{\rm
HN}$. The magnetic field at the free-expansion radius $R_{\rm HN}$
is
\begin{equation}
B=0.5\epsilon_{B,-1}^{1/2}R_{\rm HN,17}^{-1}\beta_{\rm
sh}\dot{M}_{-5}^{1/2} v_{w,3}^{-1/2} {\rm G}.
\end{equation}
From $t_{\rm acc}=t_{\rm dyn}$, the maximum energy is
\begin{equation}
\varepsilon_{\rm A, max}\simeq Z e BR_{\rm HE}\beta_{\rm sh}/\eta =
5.2\times10^{20} \eta^{-1} \left(\frac{Z}{26}\right)
\epsilon_{B,-1}^{1/2}\beta_{\rm sh}^{2}\dot{M}_{-5}^{1/2} v_{w,3}^{-1/2}
{\rm eV}.
\end{equation}
Note that the synchrotron loss of UHE nuclei in the
semi-relativistic hypernova scenario is much lower than  the
adiabatic loss, so it is not considered here.

\subsubsection{Internal shock scenario}
In the internal shock scenario, the magnetic field $B$ can be
estimated by
$B^2/8\pi=\epsilon_BU=\frac{\epsilon_B}{\epsilon_e}U_{\gamma}$ as
all the energies of electrons are almost lost into radiation, where
$U$ is the comoving thermal energy density and $U_{\gamma}=L/4\pi
R^2_{{\rm sh}}\Gamma^2c$ is the comoving gamma--ray energy density
for an internal shock radius of $R_{\rm sh}$. So,
\begin{equation}
B=\left(\frac{2\epsilon_BL}{\epsilon_ec}\right)^{1/2}\frac{1}{\Gamma
R_{{\rm sh}} } =4.30\times
10^2\epsilon_{e,-1}^{-1/2}\epsilon_{B,-1}^{1/2}L_{47}^{1/2}\Gamma_1^{-3}\delta
t_2^{-1} \rm G
\end{equation}
where $R_{\rm {sh}}=2\Gamma^2c\delta t$ has been used and $\delta t$
is the variability time.

The maximum nucleus energy is determined by equating $t_{\rm acc}$ with
the smaller one of the two cooling timescales $t_{\rm ad}$ and $t_{\rm syn}$. We
have
\begin{equation}\label{maxdyn}
\varepsilon_{A,{\rm max}}=2.0\times
10^{21}\eta^{-1}(\frac{Z}{26})\epsilon_{e,-1}^{-1/2}
\epsilon_{B,-1}^{1/2}L_{47}^{1/2}\Gamma_1^{-1}\beta_{{\rm int}} {\rm
eV}
\end{equation}
for $t_{{\rm acc}}=t_{{\rm dyn}}$, and
\begin{equation}\label{maxsyn}
\varepsilon_{A,{\rm max}}=2.3\times
10^{21}\eta^{-1/2}(\frac{A}{56})^2(\frac{Z}{26})^{-3/2}
\epsilon_{e,-1}^{1/4}\epsilon_{B,-1}^{-1/4}L_{47}^{-1/4}\Gamma_1^{5/2}\delta
t_2^{1/2}\beta_{{\rm int}} {\rm eV}
\end{equation}
for $t_{{\rm acc}}=t_{{\rm syn}}$. From Eq.~(\ref{maxdyn}), one can derive a
minimum  luminosity for LLGRBs that can accelerate nuclei to ultra
high energies, i.e.
\begin{equation}
L_{{\rm min,acc}} =2.5\times 10^{44}\varepsilon_{A,100{\rm
EeV}}^2\eta^2(\frac{Z}{26})^{-2}
\epsilon_{e,-1}\epsilon_{B,-1}^{-1}\Gamma_1^2\beta_{{\rm int}}
^{-2}{\rm ergs^{-1}}.
\end{equation}
According to Eq.~(\ref{maxsyn}), there is also a maximum luminosity for  LLGRBs
that can accelerate nuclei to ultra high energies, i.e.
\begin{equation}
L_{{\rm max,acc}}=2.8\times 10^{52}\varepsilon_{A,100{\rm
EeV}}^{-4}\eta^{-2}(\frac{A}{56})^{4}
(\frac{Z}{26})^{-6}\epsilon_{e,-1}\epsilon_{B,-1}^{-1}\Gamma_1^{10}\delta
t_2^2\beta_{{\rm int}}^{4}{\rm ergs^{-1}}
\end{equation}
The steep dependence on the bulk Lorentz factors $\Gamma$ results
from the steep dependence of the magnetic field energy on $\Gamma$
in the internal shock assumption. If the luminosity is too high, the
radiative cooling is so rapid that these nuclei can not reach
ultrahigh energies.  As the minimum luminosity adopted in Sec.2 is
higher than $L_{\rm min,acc}$, we conclude that even those missing
LLGRBs are powerful enough to accelerate UHECRs. Since the internal
shock is mildly relativistic with $\beta_{\rm int}\approx 1$,
hereafter, we drop the dependence on $\beta_{\rm int}$ in our
analytical calculation.

\subsection{Survival of UHE nuclei}
UHE nuclei will suffer from photo-disintegration and photopion loss
by interactions with low-energy photons in the sources. For UHE
nuclei, photo-disintegration is usually more important \citep{Wang08, Allard08}. Now we study whether UHE nuclei accelerated in LLGRBs can
survive in the semi-relativistic hypernova scenario and internal
shock scenario.

\subsubsection{Semi-relativistic hypernova  scenario}

There are two photon sources which could cause
photo-disintegration of heavy nuclei: one is provided by hypernova
thermal photons from radioactive elements of the hypernova ejecta,
and another is provided by the synchrotron photons from the
hypernova remnant shock. The free-expansion time for the
semi-relativistic ejecta is
\begin{equation}
t=\frac{R_{\rm HE}}{\Gamma\beta_{\rm sh}c}\simeq1.3\times10^{7} E_{k,51}
({\Gamma\beta_{\rm sh}})^{-3} \dot{M}_{-5}^{1/2} v_{w,3}^{-1/2} {\rm s}.
\end{equation}
At earlier times when the fast semi-relativistic ejecta is
decelerated, the flux from the hypernova thermal photons is expected
to be dominated.  We use the luminosity of SN1998bw as a
representative for the hypernova luminosity of thermal photons. At
time $t\sim100$ days after the burst, the optical luminosity of
SN1998bw drops to the level of about $L_{\rm HE}\sim10^{41}{\rm erg
s^{-1}}$ \citep{Sollerman02}. A nucleus of energy $E=10^{20}$ eV
interacts with target photons with a threshold energy
$\varepsilon_{\rm th}\ga0.01 ({A}/{56}) E_{20}^{-1} {\rm eV}$. A
rough estimate of the optical depth of photo-disintegration of UHE
nuclei (losing one nucleon)  due to hypernova thermal photons is
\begin{equation}
\tau\la \sigma_{\rm GDR}(\frac{L_{\rm HN}}{4\pi R_{\rm HN}^2 c
\varepsilon_{\rm HN}})(\frac{R_{\rm HN}}{\varrho}) =3\times10^{-2}
(\frac{A}{56})L_{\rm HN,41} R_{\rm HN,17}^{-1} (\frac{\varepsilon_{\rm
HN}}{ 1{\rm eV}})^{-1},
\end{equation}
where $\sigma_{{\rm GDR}}=1.45\times 10^{-27}A$cm$^2$ is the peak
cross section of the Giant Dipole Resonance (GDR), ${\varrho}\simeq
4$ is the compression ratio of the hypernova remnant shock and
$\varepsilon_{\rm HE}\simeq 1 {\rm eV}$ is the characteristic energy
of hypernova thermal photons.

\subsubsection{Internal shock scenario}
\begin{center}
{\em 1. Photon Spectrum}
\end{center}
The photon spectrum  of the prompt emissions of  GRBs can be
approximately described by a broken power-law. In the comoving frame
of the relativistic outflow, it can be described as
\begin{equation}
n(\varepsilon_{\gamma})=n_b\left \{
\begin{array}{ll}
 (\varepsilon/\varepsilon_{\gamma \rm b}^{\rm co})^{-\beta_1}, ~~~ \varepsilon<\varepsilon_{\gamma \rm b}^{\rm co},\\
 (\varepsilon/\varepsilon_{\gamma \rm b}^{\rm co})^{-\beta_2}, ~~~ \varepsilon_{\gamma \rm b}^{\rm co}< \varepsilon <\varepsilon_{{\rm max}} .\\
\end{array}
\right.
\end{equation}
where $\varepsilon_{\gamma \rm b}^{\rm co}$ is the break energy in the
comoving frame. As we did in Sec.2, we take $\beta_1=1$  and $\beta_2=2$ as typical values.
At low energy region, in the framework of the
internal shock model, there may be another spectral break, i.e. the
synchrotron self-absorption (SSA) break.

Now we estimate the SSA break energy for LLGRBs. The two
characteristic break frequencies in the synchrotron spectrum in the
comoving frame are given by
\begin{equation}
\nu_m =1.02\times 10^{12}f(p)\Gamma_1^{-3}\delta t_2^{-1}\epsilon_{e,-1}^{3/2}\epsilon_{B,-1}^{1/2}L_{47}^{1/2}(\gamma_{{\rm int}} -1)^2 \rm Hz
\end{equation}
and
\begin{equation}
\nu_c =2.37\times 10^{9}\Gamma_1^{7}\delta
t_2\epsilon_{e,-1}^{3/2}\epsilon_{B,-1}^{-3/2}L_{47}^{-3/2}(1+Y)^{-2}
\rm Hz,
\end{equation}
where $f(p)=6(p-2)/(p-1)$, $\gamma_{{\rm int}}$ is the Lorentz
factor of the internal shock, and $Y$ is the  Compton parameter for
inverse Compton loss. In the fast cooling regime ($\nu_c<\nu_m$),
$Y=\frac{-1+\sqrt{1+4\epsilon_e/\epsilon_B}}{2}\approx 0.6$ for
$\epsilon_e=\epsilon_B=0.1$. Since it is less than unity, we just
neglect its influence on the value of $\nu_c$. The SSA coefficient
in fast cooling regime can be estimated by \citep{Wu03}
\begin{equation}\label{ssa}
k_{\nu}=\left \{
\begin{array}{lll}
\frac{c_0e}{B\gamma_c^{5}}n_e\left(\frac{\nu}{\nu_c}\right)^{-5/3}, ~~~~~~~~~~~~\nu<\nu_c\\
\frac{c_0e}{B\gamma_c^{5}}n_e\left(\frac{\nu}{\nu_c}\right)^{-3}, ~~~~~~~~~~~~~~\nu_c<\nu<\nu_m\\
\frac{c_0e}{B\gamma_m^{5}}\frac{\gamma_c}{\gamma_m}n_e\left(\frac{\nu}{\nu_m}\right)^{-(p+5)/2},~ \nu_m<\nu
\end{array}
\right.
\end{equation}
where $\gamma_m$ and $\gamma_c$ is the minimum Lorentz factor and
cooling Lorentz factor, $n_e$ is the number  density of electrons
and $c_0\sim 15$ is a constant. The SSA frequency is determined by
$k_{\nu}\Delta R_{\rm co}= 1$, where $\Delta R_{\rm co}$ is the
length of the shock region in the comoving frame. $\Delta R_{\rm
co}$ is related to the column density of electrons $\Sigma$ by
\begin{equation}
\Sigma=n_e \Delta R_{\rm co}=\frac{N_{{\rm shell}}^e}{4\pi R_{\rm
int}^2}
\end{equation}
where $N_{{\rm shell}}^e=\frac{L\delta t\gamma_{{\rm int}}
}{\epsilon_e(\gamma_{{\rm int}} -1)}\frac{Z}{\Gamma m_Ac^2}$ is the
total number of electrons in the colliding shell. Finally, we get
the SSA frequency in the comoving frame, i.e.
\begin{equation}
\varepsilon_{{\rm SSA}}=0.012\Gamma_1^{-13/6}\delta t_2^{-13/18}L_{47}^{13/36}\rm eV
\end{equation}
for $\epsilon_e=0.1$, $\epsilon_B=0.1$, $\gamma_{{\rm int}} =2$ and $p=2.2$.

Since $\nu_{{\rm SSA}}>\nu_m$, the SSA coefficient is $\propto
\varepsilon^{-3.6}$ for $p=2.2$ according to Eq.~(\ref{ssa}),  hence the
spectral index below the SSA frequency is $-2.6$. Thus, the photon
spectrum in the comoving frame is
\begin{equation}
n(\varepsilon)=n_b\left \{
\begin{array}{lll}
 (\varepsilon_{{\rm SSA}}/\varepsilon_{\gamma \rm b}^{\rm co})^{-1}(\varepsilon/\varepsilon_{{\rm SSA}})^{2.6}, \varepsilon<\varepsilon_{{\rm SSA}}\\
 (\varepsilon/\varepsilon_{\gamma \rm b}^{\rm co})^{-1}, ~~~~ \varepsilon_{{\rm SSA}}<\varepsilon<\varepsilon_{\gamma \rm b}^{\rm co},\\
 (\varepsilon/\varepsilon_{\gamma \rm b}^{\rm co})^{-2}, ~~~ \varepsilon_{\gamma \rm b}^{\rm co}<\varepsilon<\varepsilon_{{\rm max}} .\\
\end{array}
\right.
\end{equation}

\begin{center}
{\em 2. Photodisintegration }
\end{center}
For an UHE nucleus with energy $\varepsilon_A$ in the observer
frame, the fractional photo-disintegration rate is
\begin{equation}\label{tau_num}
t_{{\rm dis,f}}^{-1}(\gamma_A)=\frac{1}{A}|\frac{dA}{dt}|=\frac{c}{2\gamma_A^2}\int_{\epsilon_{{\rm
th}}}^\infty d\epsilon\sigma_{{\rm dis}} (\epsilon)\epsilon
\int_{\epsilon/2\gamma_A}^{\infty}dx x^{-2}n(x)
\end{equation}
where $\gamma_A=\varepsilon_A/\Gamma m_Ac^2=1.9\times
10^8(A/56)^{-1}\Gamma_1^{-1}\varepsilon_{A,{\rm 100EeV}}$,
$\sigma_{{\rm dis}}(\epsilon)$ is the photo-disintegration total
cross section, $\epsilon$ is the photon energy in the nucleus rest
frame and $\epsilon_{{\rm th}}$ is the threshold energy. This rate
can be calculated numerically with the cross section given in
\citet{Puget76}. But as an estimate  we can approximate the cross
section mainly contributed by the giant dipole resonance (GDR) and
show the numerical results later. The peak cross section due to GDR
is $\sigma_{{\rm GDR}}=1.45\times 10^{-27}A$cm$^2$ and
$\epsilon_{{\rm GDR}}=42.65A^{-0.21}$MeV (for $A>4$) with a width
$\Delta_{{\rm GDR}}=8$MeV. We find that the energy of the photons
that interact with UHE nuclei via GDR resonance is about
$\varepsilon_{{\rm GDR}}^{{\rm co}}\sim
0.2\Gamma_1(A/56)^{0.79}(\varepsilon_{A,50{\rm EeV}})^{-1}$eV in the
comoving frame, which is larger than the SSA frequency, so the SSA
process in the low-energy photons will have little influence on
photo-disintegration process for these representative parameters.
Approximating the integral by the contribution from the resonance,
the fractional photo-disintegration rate  is
\begin{equation}
t_{{\rm dis,f}} ^{-1}=\frac{U_{\gamma}}{4\varepsilon_{\gamma
b}^{\rm co}}\frac{c\sigma_{{\rm GDR}}\Delta_{{\rm GDR}}}{A\epsilon_{{\rm
GDR}}}\left \{
\begin{array}{ll}
(\varepsilon_A/\varepsilon_{A\rm b})^{\beta_1 -1}~~~\varepsilon_A>\varepsilon_{A\rm b},\\
(\varepsilon_A/\varepsilon_{A\rm b})^{\beta_2 -1}~~~\varepsilon<\varepsilon_{A\rm b},\\
\end{array}
\right.
\end{equation}
where $U_{\gamma}\approx n_{\rm b}(\varepsilon_{\rm b}^{\rm
co})^2[1+{\rm ln}(\varepsilon_{350{\rm keV}}/\varepsilon_{\gamma \rm
b})]\approx 4n_{\rm b}(\varepsilon_{\gamma \rm b}^{\rm co})^2$ is
the energy density of photons in the comoving frame in the energy
window of \textit{Swift}/BAT  and $\varepsilon_{A\rm
b}=0.5\epsilon_{{\rm GDR}}m_Ac^2\Gamma^2/\varepsilon_{\gamma \rm b}
\simeq 1.5\times 10^{15}A\Gamma_1^2\varepsilon_{\gamma \rm b,{\rm
10keV}}^{-1}$eV is the nuclei break energy in the observer frame.
For UHE nuclei, we have $\varepsilon_A\gg \varepsilon_{A\rm b}$,
then the effective optical depth for photo-disintegration is
\begin{equation}\label{tau}
\tau_{{\rm dis,f}}=\frac{t_{{\rm dyn}}}{t_{{\rm dis,
f}}}=0.043\frac{L_{47}(A/56)^{0.21}}{\Gamma_1^4\delta t_2
\varepsilon_{\gamma \rm b,{\rm 10keV}}}
\end{equation}
for $\beta_1=1$. Thus we can see that a larger break energy, a
larger bulk Lorentz factor  or a smaller luminosity  will be
favourable for the survival of UHE heavy nuclei. We note that these
quantities in GRBs may have inherent correlations.

This motivates us to study in which kind of bursts UHE nuclei can survive,
especially whether UHE nuclei can survive in those missing, dim LLGRBs, when
these inherent correlations are taken into account. For the break energy
$\varepsilon_{\gamma \rm b}$, we still assume that the Yonetoku relation holds for
LLGRBs, i.e. $\varepsilon_{\gamma \rm b} \propto L^{1/2}$. As to
$\Gamma$, assuming the break energy of the spectrum in the comoving
frame are constant for all LLGRBs\footnote{Given that LLGRBs have
lower break energies than HLGRBs, and on the other hand, LLGRBs are
suggested to arise from less relativistic jets \cite{Soderberg06b,
Toma07}, the intrinsic range of the break energy in the comoving
frame could be small.}, we have $\Gamma\propto \varepsilon_{\gamma
\rm b}$ and hence $\Gamma\propto L^{1/2}$. The variability time
scale $\delta t$ may depend on the central engine activity  and we
simply fix its value to 100s. Under these assumptions, we have
$\tau_{{\rm dis,f}}\propto L^{-3/2}$, and consequently, UHE heavy
nuclei can survive more easily in LLGRBs with relatively high
luminosities. The main reason for this is that LLGRBs with lower
luminosity may have smaller internal shock radii for a fixed $\delta
t$ and hence a higher density of target photons. On the other hand,
if we assume that $\Gamma$ is a constant for LLGRBs, we will obtain
$\tau_{{\rm dis,f}}\propto L^{1/2}$. In this case, UHE heavy nuclei
can survive more easily in LLGRBs with relatively low luminosities.

The SSA break may affect the photo-disintegration rate when
$\varepsilon_{{\rm GDR}}^{{\rm co}}<\varepsilon_{{\rm SSA}}$,
because the number of photons that can interact with the heavy
nuclei  via GDR is considerably reduced by the SSA process. This is
applicable  for  nucleus at the highest energy (i.e. typically above
a few times $10^{20}$eV). In the SSA regime, we have
\begin{equation}
t_{{\rm dis,f}} ^{-1}\approx \frac{U_{\gamma}}{12.8\varepsilon_{\gamma \rm
b}^{\rm co}}\frac{c\epsilon_{{\rm GDR}} \sigma_{{\rm GDR}}\Delta_{{\rm
GDR}}}{A\gamma_A^2\varepsilon_{{\rm SSA}}^2}
\end{equation}
The effective photo-disintegration  optical depth is
\begin{equation}
\tau_{{\rm dis,f}}(\varepsilon_{A})=0.039
\frac{L_{47}^{5/18}(A/56)^{1.79}}{\Gamma_1^{-7/3}\delta t_{2}^{-4/9}
\varepsilon_{\gamma \rm b,{\rm 10keV}}(\varepsilon_{A,500{\rm
EeV}})^2}.
\end{equation}
If we assume $\varepsilon_{\gamma \rm b}\propto L^{1/2}$ and
$\Gamma\propto \varepsilon_{\gamma \rm b}$, we will have
$\tau_{{\rm dis,f}}\propto L^{17/18}$, hence UHE heavy nuclei can
survive from photo-disintegration more easily in LLGRBs with
relatively low luminosities, if the energy of the effective target
photons falls below the SSA break energy.

\begin{figure}
\resizebox{\hsize}{!}{\includegraphics{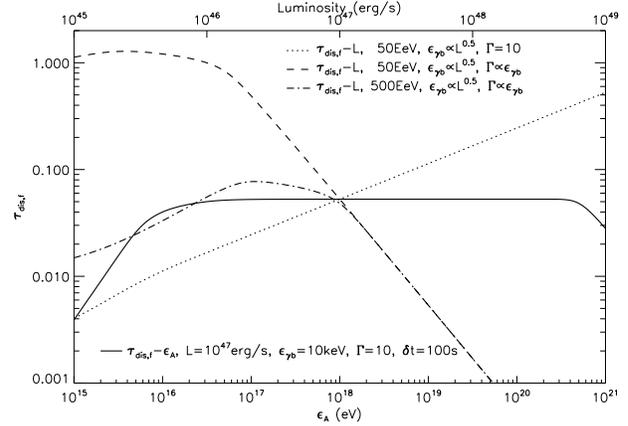}} \caption{{ The
fractional optical depth for photodisintegration of iron nuclei in
the internal shock scenario. It is a function of the energy of
nuclei (the solid line) and is a function of the luminosity of
LLGRBs (dashed lines).} The parameters used in the calculation are
given in the figure for each lines. See text for more details.}
\label{fig3}
\end{figure}

In Fig.3, we show the effective fractional photo-disintegration optical depth
$\tau_{{\rm dis,f}}$ for iron nucleus, calculated according to
Eq.~(\ref{tau_num}) and using the cross section described by the
Lorentzian form in the energy range $\epsilon_{\rm th}<\epsilon<30$
MeV and a flat cross section for multi-nucleon loss in the energy
range $30 {\rm MeV}<\epsilon<150{\rm MeV}$ \citep{Puget76,
Karakula93}. From the solid line, one can see that   $\tau_{{\rm
dis,f}}$ is almost independent of the energy of  nucleus
$\varepsilon_A$ when $\varepsilon_A>\varepsilon_{A\rm b}$ (see the
approximate analytic estimate in Eq.~(\ref{tau})). When the energy
of the nucleus becomes larger than several times $10^{20}$ eV, the nucleus
starts to interact with target photons with energies below the SSA
break  so that the effective optical depth decreases with the energy
of nucleus.  Fig.~3 (the dotted, dashed and dash-dotted lines) also
describes how $\tau_{{\rm dis,f}}$ of an UHE nucleus with a fixed
energy depends on the luminosity of LLGRB. We can see that under
the assumptions of $\varepsilon_{\gamma \rm b}\propto L^{1/2}$ and
$\Gamma\propto \varepsilon_{\gamma \rm b}$, UHE nuclei tend to
survive in LLGRB with relative high luminosity.  The relation
between $\tau_{{\rm dis,f}}$ and $\varepsilon_A$ in the assumption
that $\Gamma$ is a constant for all LLGRBs is also presented in
Fig.~3 for comparison. In this case, UHE nuclei survive more easily
in LLGRB with lower luminosity.

\begin{figure}
\resizebox{\hsize}{!}{\includegraphics{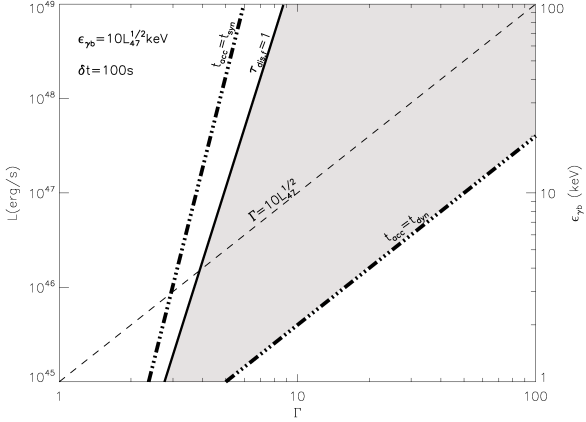}} \caption{{ The
parameter space of luminosity $L$ and Lorentz factor $\Gamma$ in the
internal shock scenario.} The hatched area is the parameter space of
$L$ and $\Gamma$ of LLGRBs in which iron nuclei of energy 50 EeV can
both be accelerated and survive in the sources. The thick solid line
represents the condition $\tau_{\rm dis,f}=1$. The area between the
two thick dash-dotted lines represent the parameter space in which
iron nuclei can be accelerated to 50 EeV. Also shown is the
Yonetoku's relation for GRBs (the thin dashed line).} \label{fig4}
\end{figure}

In Fig.~4, we show the parameter space (denoted by the hatched area)
of $L$ and $\Gamma$ in which  iron nucleus can be accelerated to
$50$ EeV and can escape from the source with most of its initial
nucleons preserved meanwhile.  It shows that LLGRBs with larger $L$
tend to satisfy both constraints in the internal shock scenario.
Nevertheless,  $\tau_{\rm dis,f}\simeq 1$ occurs at $L\simeq
10^{46}\rm ergs^{-1}$, which is sufficiently low that most LLGRBs we
are concerning are within the safety limit.

\section{Neutrino background from LLGRBs}
UHE nuclei will interact with low-energy photons in the sources and
produce high-energy neutrinos.  Hence neutrino detection is a useful
tool for probing the acceleration of UHECRs in  sources. In the
semi-relativistic hypernova scenario, the production of neutrinos is
mainly through the photomeson interactions of UHE nuclei with
hypernova thermal photons \citep{Wang07a}. It has been found that
the diffuse neutrino flux contributed by semi-relativistic hypernova
is too low to be detected by cubic  kilometer  detectors such as
Icecube \citep{Wang07a}.

In internal shock scenario, the production of neutrinos is mainly
through the photomeson interactions of UHE nuclei with low-energy
prompt photons in LLGRBs. The fractional energy loss rate of a
nucleus with energy $\varepsilon_A$ in the comoving  frame due to
photomeson productions is
\begin{equation}
\begin{split}
t_{{\rm mes}}^{-1}(\gamma_A)&=\frac{1}{\gamma_A}|\frac{d\gamma_A}{dt}|_{{\rm mes}}\\
&=\frac{c}{2\gamma_A^2}\int_{\epsilon_{{\rm th}}}^\infty
d\epsilon\sigma_{{\rm mes}}(\epsilon)\xi_A(\epsilon)\epsilon
\int_{\epsilon/2\gamma_A}^{\infty}dx x^{-2}n(x)
\end{split}
\end{equation}
where $\epsilon$ is the photon energy in the rest frame of the
nucleus,  $\epsilon_{{\rm th}}=0.15$GeV is the threshold photon
energy for photomeson interaction, $\sigma_{{\rm mes}}(\epsilon)$ is
the cross section of photomeson production  and $\xi_A(\epsilon)$ is
the average fraction of energy lost to secondary pions. Above the
threshold energy, the main contribution of pion production is due to
the $\Delta$ resonance at $\epsilon_{\Delta} = 0.34$GeV, for which
the cross section can be approximately by a Lorentzian form (e.g.
\citealt{Mucke00}). Neglecting the nuclear shadowing effect in
photoproduction \citep{Michalowski77}, the peak cross section is
$\sigma_{{\rm \Delta}}\simeq \sigma_{p\gamma}A\simeq 4.1\times
10^{-28}A$cm$^2$ \citep{Anchordoqui08, Murase08}, where
$\sigma_{p\gamma}$ is the peak cross section for protons. As
$\xi_A(\epsilon)\sim \xi_p(\epsilon)/A$, $\sigma_{{\rm
\Delta}}\xi_A$ is independent of $A$. The energy of the photons that
interact with UHE nuclei via $\Delta$ resonance is
$\varepsilon_{\Delta}^{{\rm co}}\sim
3.8\Gamma_1(A/56)(\varepsilon_{A,50{\rm EeV}})^{-1}$eV in the
comoving frame . Since this energy is much larger than the SSA
break energy, the SSA process  has little effect on the
photomeson process, even for the nucleus with energy of a few times
$10^{20}$eV. Approximating the integral by the contribution from the
resonance, the total fraction of energy lost by nuclei to pions  is
\begin{equation}
f_{\pi}(\varepsilon_A)=\frac{t_{{\rm dyn}}}{t_{\rm mes}}=
\frac{R_{{\rm int}} }{\Gamma ct_{\rm mes}}\simeq
0.004\frac{L_{47}}{\varepsilon_{\gamma \rm b,{\rm 10keV}}\Gamma_1 ^4\delta t_2} .
\end{equation}
So $f_{\pi}(\varepsilon_A)$ depends on the inherent relations among
$\Gamma$, $\varepsilon_{\gamma \rm b}$ and $L$. For $\varepsilon_{\gamma
b} \propto L^{1/2}$ and $\Gamma\propto\varepsilon_{\gamma \rm b}$, we
have $f_{\pi}(\varepsilon_A)\propto L^{-3/2}$. But in the case that
the bulk Lorentz factor is a constant with $\Gamma=10$,
$f_{\pi}(\varepsilon_A)\propto L^{1/2}$.  For LLGRBs with luminosity
in the range of $L\sim 10^{46-49} {\rm erg s^{-1}}$ , $f_{\pi}\ll 1$
for both cases, which indicates that  most heavy nuclei suffer from
negligible energy loss due to photomeson production process.

The diffuse neutrino flux contributed by all LLGRBs can be obtained
through the integration over the whole luminosity range  of LLGRBs,
which is
\begin{equation}
\varepsilon_{\nu}^2\frac{dN_{\nu}}{d\varepsilon_{\nu}}(\varepsilon_{\nu})=
\frac{1}{4}\frac{c}{4\pi H_0} f_z\int_{L_{{\rm
min}}(\varepsilon_A)}^{L_{{\rm max}}}{\rm
min}[1,f_{\pi}(L)]\varepsilon_A^2\frac{dN_A}{d\varepsilon_A}\zeta_{\pi}\frac{dN}{dL}dL
\end{equation}
where $f_z\sim3$ is the correction factor for the contribution from
high redshift sources and $f_{\pi}$  is the suppression factor on
the neutrino flux due to pion cooling \citep{Rachen98, Razzaque04,
Wang09}. Since the synchrotron cooling dominates the cooling
process, we have $\zeta_{\pi}=\tau_{\pi}^{-1}/(t_{\pi,\rm
syn}^{-1}+\tau_{\pi}^{-1})$, where $t_{\rm
syn,\pi}=12L_{47}^{-1}\Gamma_1^7\delta t_2^2(\varepsilon_{\pi,1\rm
EeV})^{-1}$s is the synchrotron cooling time and
$\tau_{\pi}=2.6\times 10^{-8}\gamma_{\pi}=186\varepsilon_{\pi,1\rm
EeV}$s is the lifetime of pions. The lower limit of the integral
$L_{{\rm min}} (\varepsilon_A)$ is given in Table.~1. The results
obtained numerically are shown in Fig.~5 for the two LFs. It shows
that only in the case of high local rate of LLGRBs, there is a
chance that diffuse neutrinos from LLGRBs could be detected by cubic
kilometer detectors such as Icecube.

\begin{figure}
\resizebox{\hsize}{!}{\includegraphics{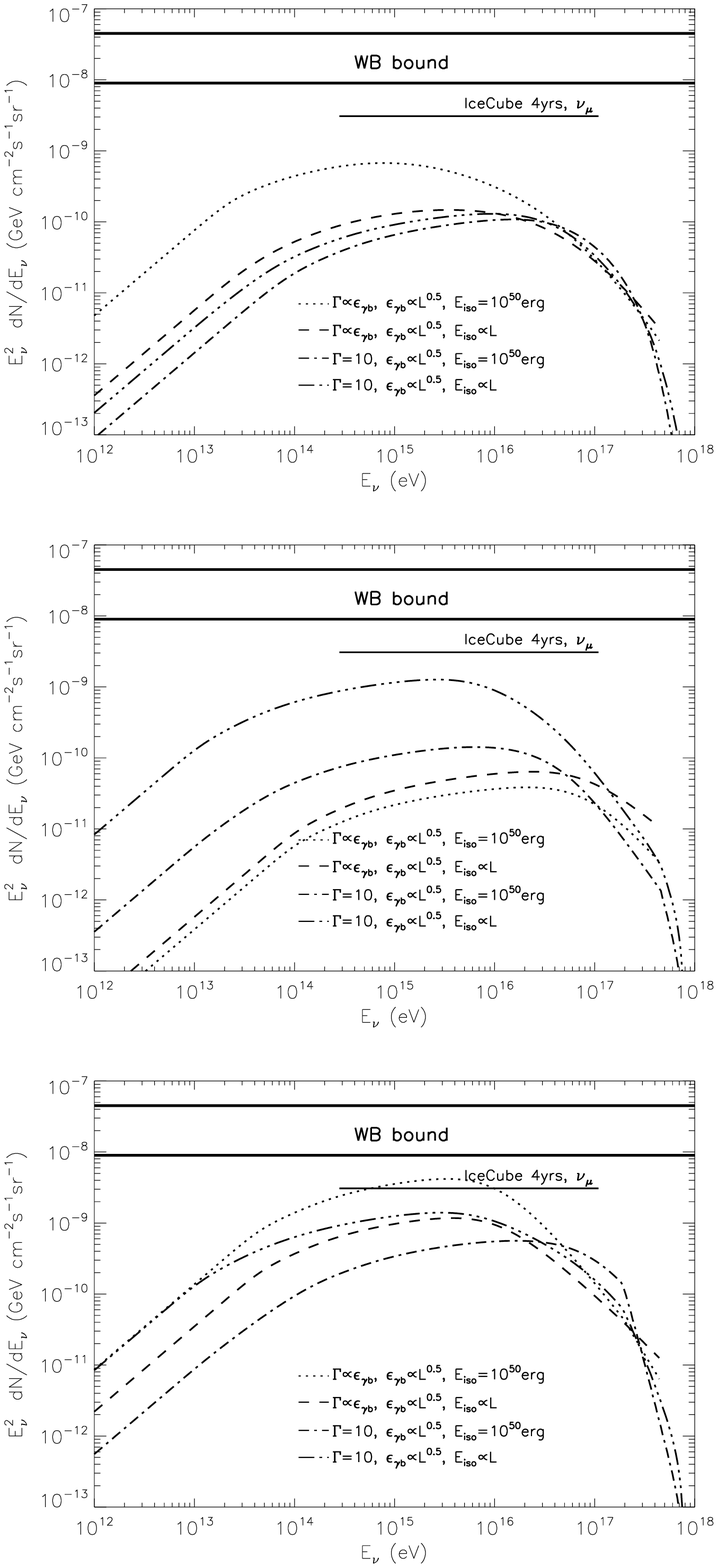}} \caption{The diffuse neutrino
background from LLGRBs.  The top panel is for the LF in Liang et al.
(2007), whereas the middle and the bottom ones are for the LF in Dai
(2009), but for a local rate of $\rho_0=200$Gpc$^{-3}$yr$^{-1}$ and
$\rho_0=2000$Gpc$^{-3}$yr$^{-1}$ respectively. The reference values
of $\varepsilon_{\gamma \rm b}$, $\Gamma$ and $E_{\gamma,\rm iso}$ in the
scalings are 10keV, 10 and $10^{50}$erg respectively for a LLGRB
with luminosity of $10^{47}\rm ergs^{-1}$.  { We also plot the IceCube
sensitivity to diffuse neutrino fluxes in 4 years
operation \citep{Spering08}} and the Waxman-Bahcall bounds (Waxman
\& Bahcall 1999).}\label{fig5}
\end{figure}

\section{Discussions and Conclusions}
The origin of heavy or intermediate-mass nuclei in LLGRBs remains to
be an open issue. In the hypernova remnant scenario,   cosmic ray
particles originate from the material swept-up by the
semi-relativistic hypernova shock front. The progenitors of
hypernovae are thought to be Wolf-Rayet stars, as the spectral type
of the discovered supernovae in these events is typically Ic. These
stars are stripped of their hydrogen envelope and sometimes even the
helium envelope, so the circum-stellar wind is rich of heavy
elements, such as C and O. Thus, heavy or intermediate-mass UHE
nuclei may naturally originate from the element-enriched stellar
wind of Wolf-Rayet stars in the hypernova scenario. In WC type
Wolf-Rayet stars, the C abundance is $X_{\rm C}\simeq20\%-55\%$ (by
mass) and the O abundance is $X_{\rm O}\simeq5\%-10\%$ \citep{Crowther07}.
In WO type Wolf-Rayet stars, O and C abundances are even
higher, with $X_{\rm C}\simeq40\%-55\%$ and $X_{\rm
O}\simeq15\%-25\%$ (e.g. \citealt{Kingsburgh95}). The abundances of
these elements are clearly much higher than the solar values. In the
internal shock scenario for LLGRBs, the origin of nuclei is,
however, quite uncertain. The uncertainty lies in whether there are
nuclei entrained into the relativistic jet during the formation of
the jet out of the collapsing core  of massive stars. The temperature
of the accretion disk resulted from the collapsing core needs to be
sufficiently low that nuclei in the disk will not be
photo-disintegrated by the disk thermal photons. This condition may
be satisfied if the accretion rate for LLGRBs is sufficiently low.
Another possibility for the origin of nuclei in the internal shock
scenario is that nuclei present in the surrounding stellar envelope
are entrained into the jet while jet is propagating through the
stellar envelope. The validity of this possibility needs a detailed
numerical simulation of the jet propagation in LLGRBs.

In this paper, we have suggested that many dim local low luminosity GRBs
may be missed by \textit{Fermi}/GBM and these missing LLGRBs could
make a dominant contribution to observed flux of UHECRs. If true, it
can relieve partly the  discrepancy  between the energy production
rate in UHECRs  and that in gamma-ray photons recorded by
\textit{Fermi}/GBM, as raised by \citet{Eichler10}. We first
calculate the energy production rate in gamma--ray photons by
LLGRBs as a separate population from high-luminosity GRBs and
estimate how large  those missing LLGRBs contribute to the total
gamma--ray energy production. In the calculation, we  take two
different LFs for LLGRBs that were proposed in the literatures. We
find that the gamma-ray energy production rate by LLGRBs for both
LFs are one to two orders of magnitude larger than that estimated
by \citet{Eichler10}. The missing part of the energy production rate
in gamma-rays could account for a fair proportion or even the vast
majority of the total one and hence could be much larger than the
observable part. It should be noted that our results depend on many
assumptions, such as the form of LF, the spectrum of LLGRBs and some
inherent relations among parameters of LLGRBs (e.g. $E_{\gamma}-L$
relation, $\varepsilon_{\gamma \rm b}-L$ relation). To get a more
accurate results, further studies in these relations with larger
statistics are needed.

There are two scenarios for the UHECR acceleration in LLGRBs, one is
the semi-relativistic hypernova scenario where UHECRs are
accelerated by the semi-relativistic hypernova shock expanding into
the circum-stellar wind and another is the internal shock scenario
where UHECRs are accelerated by internal shocks within the variable
relativistic outflow. We find that, in both scenarios, only
intermediate-mass or heavy nuclei could be accelerated to energies
above $10^{20}$ eV, which is consistent with the recent findings of
heavy UHECR composition by PAO.  We have also studied whether heavy
nuclei could survive from photo-disintegration by low-energy photons
in the sources. In the semi-relativistic hypernova scenario, UHE
nuclei can survive easily as the large size of the acceleration
region leads to a low density of target photons. On the other hand,
the survival probability of UHE nuclei in internal shock scenario
depends on the inherent relations among the quantities of LLGRBs,
such as $\varepsilon_{\gamma \rm b}$, $\Gamma$ and $L$. In spite of
the assumptions in these relations,  we find that in the luminosity
range of LLGRBs that we are concerning, UHE heavy nuclei can retain
most of their nucleons before escaping from the sources. Finally,
the accompanying neutrino flux is calculated. The energy loss
through photopion production is negligible for UHE nuclei and hence
the diffuse neutrino flux from LLGRBs can hardly be detected by
current detectors, except in the case of a high local rate of
LLGRBs.

Our calculations indicate that  local LLGRBs remain viable sources
of UHECRs in terms of the energy production rate. Future GRB
detectors, such as \textit{EXIST}/HET, could cover a wider burst
trigger band (5-600keV) with an advanced sensitivity of $1.5\times
10^{-9}\rm ergcm^{-2}s^{-1}$ \citep{Imerito08}.   \textit{EXIST}/HET
will increase the LLGRB detection rate by a factor of tens compared
to \textit{Swift}/BAT and \textit{Fermi}/GBM \citep{Imerito08}, thus
much more LLGRBs could be detected. It would help us to learn more
about the nature of LLGRBs and examine more carefully the connection
between LLGRBs and UHECRs in future.

\section*{Acknowledgements}
This work is supported by the NSFC under grants
10873009, 10973008 and 11033002, the 973 program under grants
2009CB824800 and 2007CB815404, the Program for New Century Excellent
Talents  in University, the Qing Lan Project and the Fok Ying Tung
Education Foundation.

\end{document}